# 4MOST Consortium Survey 4: Milky Way Disc and Bulge High-Resolution Survey (4MIDABLE-HR)

Thomas Bensby[1]
Maria Bergemann[2]
Jan Rybizki[2]
Bertrand Lemasle[3]
Louise Howes[1]
Mikhail Kovalev[2]
Oscar Agertz[1]
Martin Asplund[4]
Paul Barklem[5]
Chiara Battistini[6]
Luca Casagrande[4]
Cristina Chiappini[7]
Ross Church[1]
Sofia Feltzing[1]
Dominic Ford[1]
Ortwin Gerhard[8]
Iryna Kushniruk[1]
Georges Kordopatis[9]
Karin Lind[2,5]
Ivan Minchev[7]
Paul McMillan[1]
Hans-Walter Rix[2]
Nils Ryde[1]
Gregor Traven[1]

[1] Lund Observatory, Lund University, Sweden
[2] Max-Planck-Institut für Astronomie, Heidelberg, Germany
[3] Zentrum für Astronomie der Universität Heidelberg / Astronomisches Rechen-Institut, Germany
[4] Research School of Astronomy & Astrophysics, Australian National University, Canberra, Australia
[5] Department of Physics and Astronomy, Uppsala universitet, Sweden
[6] Zentrum für Astronomie der Universität Heidelberg / Landessternwarte, Germany
[7] Leibniz-Institut für Astrophysik Potsdam (AIP), Germany
[8] Max-Planck-Institut für extraterrestrische Physik, Garching, Germany
[9] Observatoire de la Côte d'Azur, Nice, France

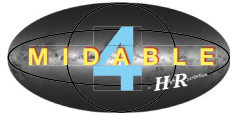

The signatures of the formation and evolution of a galaxy are imprinted in its stars. Their velocities, ages, and chemical compositions present major constraints on models of galaxy formation, and on various processes such as the gas inflows and outflows, the accretion of cold gas, radial migration, and the variability of star formation activity. Understanding the evolution of the Milky Way requires large observational datasets of stars via which these quantities can be determined accurately. This is the science driver of the 4MOST MIlky way Disc And BuLgE High-Resolution (4MIDABLE-HR) survey: to obtain high-resolution spectra at $R \sim 20\,000$ and to provide detailed elemental abundances for large samples of stars in the Galactic disc and bulge. High data quality will allow us to provide accurate spectroscopic diagnostics of two million stellar spectra: precise radial velocities; rotation; abundances of many elements, including those that are currently only accessible in the optical, such as Li, s-, and r-process; and multi-epoch spectra for a sub-sample of stars. Synergies with complementary missions like Gaia and TESS will provide masses, stellar ages and multiplicity, forming a multi-dimensional dataset that will allow us to explore and constrain the origin and structure of the Milky Way.

## Scientific context

One of the key questions in astrophysics is to understand the assembly history and evolution of the Milky Way, as our understanding of galaxy formation in the Universe — be it using observations or models — is only as good as our knowledge of our own Galaxy. This requires large datasets that provide reliable physical characterisation of stars across the full Hertzsprung-Russell diagram, including precise velocities, ages, multiplicity, rotation, and elemental abundances for all nucleosynthesis channels. This is the main goal of the 4MIDABLE-HR survey.

It has been established from star counts that the Galactic disc has two components, the thin and thick discs (Gilmore & Reid, 1983). This morphological dichotomy could be related to the distinct elemental abundance trends observed in the solar neighbourhood, as well as in the inner and outer regions of the Galactic disc (for example, Bensby et al., 2014; Hayden et al., 2015). Other studies, however, either find no clear separation in the local volume of about 1 kpc or place it at a different location in elemental abundance space (for example, Bergemann et al., 2014). Cosmological simulations of galaxy formation suggest that galaxies like the Milky Way may experience various evolutionary histories, with or without multi-model structures arising in the elemental abundance plane (for example, Grand et al., 2018). On the other hand, radial migration of stars (for example, Schönrich & Binney, 2009) might have a strong impact on the disc morphology and on the observable distributions, such as the age-metallicity relationships and spatial distribution of elemental abundances. The mere presence and the size of the elemental abundance trend gap, its position in the age and abundance planes, and its relationship to stellar motions, are decisive constraints on the models of secular evolution and heating of the disc, as well as on the gas accretion and merger history of the Milky Way (for example, Rix & Bovy, 2013).

Further complexities are connected to our present understanding of the inner Galaxy. It has been established that it contains a boxy peanut-shaped bar that impacts the dynamical properties of the disc. Spectroscopic observations suggest that the bulge comprises a very complex pattern in the age-abundance plane, ranging from metal-rich young α-poor to metal-poor and very old α-rich stellar populations (for example, Ness et al., 2013; Bensby et al., 2017). It is not clear whether the bulge and the α-rich component of the disc share the same origin, or if this similarity is the consequence of high star formation efficiencies, but separate formation scenarios. Analysis of the photometric colour-magnitude diagrams suggests that the bulge has temporal properties (for example, Renzini et al., 2018) different from those revealed by spectroscopic observations of microlensed dwarf stars (Bensby et al., 2017). How much room is there for a classical bulge component made by mergers (for example, Barbuy et al., 2018)? A major challenge is to explain how all these constituents fit together and how they link to the chemo-dynamical structure of the Galactic disc and halo.

Disentangling all of these building blocks and the role that the different physical ingredients play in the formation of the Milky Way is hard. It requires a densely sampled homogeneous sample of stars with accurate elemental abundances, kinematics, and ages, requiring not only elemental abundances of α-elements, but also of all major nuclear channels: Li, C, N, O, the α-elements, the iron-group, and the neutron-capture r- and s-process elements. Only such a dataset will provide the requisite combination of constraints on the gas flows, star formation, and detailed chemical evolution of the Milky Way.

These are the observed quantities that 4MIDABLE-HR aims to provide for about two million stars in the Milky Way disc and bulge; stellar motions will be derived by combining accurate radial velocities from our spectra with proper motions and parallaxes from Gaia and special sub-surveys in 4MIDABLE-HR will address Cepheids, deep fields in the bulge, and the deep thick disc/halo field in the 4MOST Wide Area VISTA Extragalactic Survey viewing zone. These data will provide a unique treasure trove for high-precision stellar physics and will support the exploration of the Galaxy's evolution through: (i) a detailed investigation of the disc sub-structure throughout the Milky Way; (ii) quantifying the role of secular processes, such as the strength of radial migration, resolved by time and galactocentric radius; (iii) unveiling the stellar population content and its chemo-dynamical characteristics of the Galactic bulge; (iv) constraining the formation time and growth rate of the Galactic bar. In contrast to 4MIDABLE-LR (Chiappini et al., p. 30), the 4MIDABLE-HR Survey will focus on brighter stars and will aim to obtain spectra with sufficient quality and as many complementary diagnostics as possible from asteroseismology and astrometry, to provide a baseline against which the fundamental stellar parameters of all other 4MOST Galactic surveys can be assessed.

4MIDABLE-HR is unique amongst the other recent, ongoing, and planned high-resolution spectroscopic surveys in several respects. It is the largest optical high-resolution survey in terms of the number of targets, photometric depth, and survey area. Gaia-ESO is slightly deeper, but has patchy sky coverage, lower signal-to-noise ratio (S/N), and also many fewer stars. The WEAVE survey will not target the Galactic bulge region or areas close to the Galactic plane.

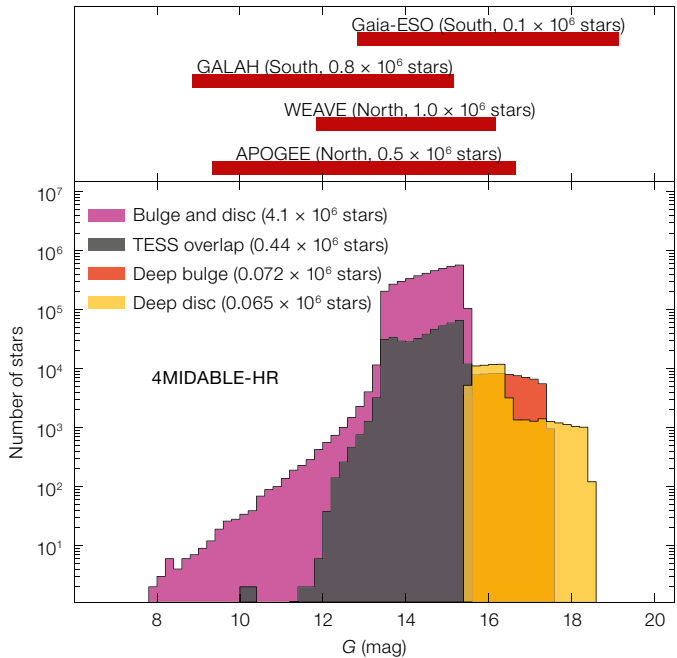

Figure 1. Magnitude distributions of our main bulge and disc sample and our two deep fields, one towards the bulge and one towards the 4MOST WAVES fields. The horizontal lines in the upper panel mark the magnitude ranges of selected high-resolution spectroscopic surveys, as indicated.

The focus of GALactic Archaeology with the HERMES spectrograph (GALAH) is mainly on the brighter targets, $V \lesssim 14$ magnitudes, and it does not probe such great distances into the Galaxy as we will do. The Apache Point Observatory Galactic Evolution Experiment (APOGEE) is an infrared survey and will not analyse neutron-capture elements — also, by probing to magnitudes $H < 14$, it will not have the depth of 4MIDABLE-HR. Figure 1 shows a comparison between 4MIDABLE-HR and these surveys.

Our spectra will also allow transformational studies of stellar physics thanks to the large sample of stars at all evolutionary stages, from the main-sequence, through the red bump to the core He-burning and AGB phases, and including pulsators like Cepheids. For 15% of our stars, asteroseismic information from the TESS exoplanet mission will be available. This information, combined with our spectroscopic characterisation, will not only allow us to put new constraints on the interior structure of stars, but also help to constrain the masses and ages of those stars and to assist planet-search programmes with characterising the planet-hosting stars. A significant fraction of stars are binaries, or even triples, allowing new constraints on the evolution and interaction of objects in multiple systems, especially when combined with binarity information from Gaia astrometric data and photometric variability from Optical Gravitational Lensing Experiment (OGLE) and Large Synoptic Survey Telescope data.

### Specific scientific goals

– What is the growth history of the Milky Way?

Our survey will give us direct information on the kinematics and detailed abundances in the outer disc and in the inner halo, which will help to constrain the masses and time of infall of the merging satellites. The signatures are very weak and require precise elemental abundances. In addition to the clustering in elemental abundance space, the merger history can be traced by physical over-densities and streams in the disc. Also, the velocity distribution in the vicinity of the Sun is not smooth, but contains a lot of kinematical over-densities (for example, Kushniruk et al., 2017). These could be due to dynamical resonances with the Galactic bar or spiral arms, dispersed open clusters, remnants of merged satellite galaxies, or even "ringing" signatures of a satellite galaxy analogous to the outer disc structures. The true nature of such structures can be revealed through detailed comparisons of the elemental abundances in the stars in the structures with measurements in the foreground

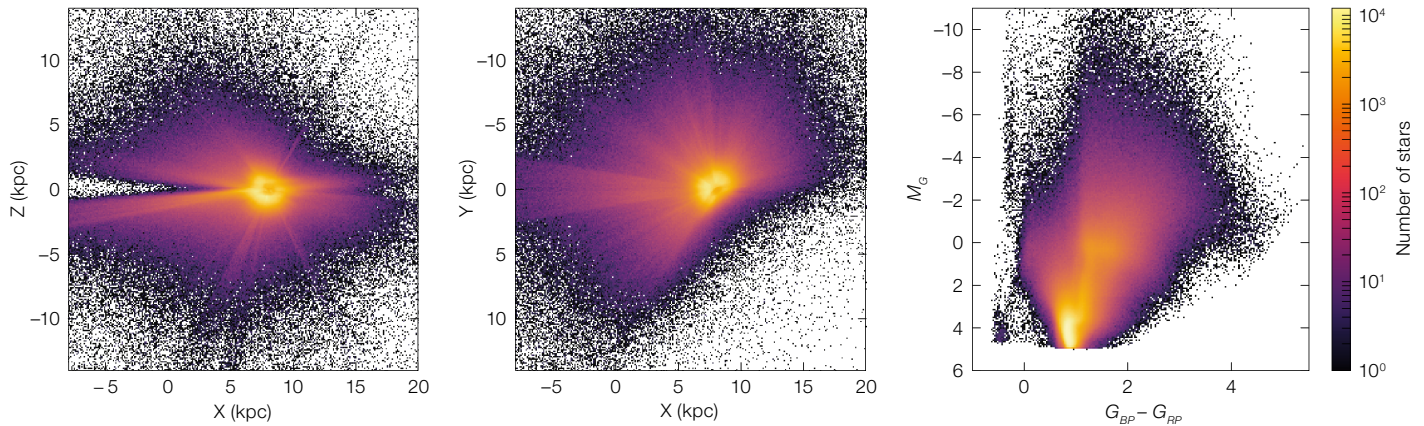

Figure 2. The left-most and middle plots show the distribution of the targets in the Galactic Cartesian x-y-z coordinate system of our input catalogue (with a bin size of 100 × 100 pc). The right-most plot shows a colour-magnitude diagram, using the Gaia G magnitudes and $G_{BP}$-$G_{RP}$ colours (not corrected for extinction).

and background stars in the Milky Way (Bergemann et al., 2018).

– Why does the Milky Way disc have such distinct abundance structure patterns? Are the thin and thick Galactic discs separate or are they a manifestation of observational biases? Could it be that the Milky Way only has one disc, but that radial migration has rearranged stars to create a sub-structure in the chemical and kinematic spaces? Can we confirm or rule out the hiatus in star formation between the two components? To address these important questions, we need large and unbiased samples of stars beyond the solar neighbourhood that cover a broad range of distances, from the inner to the outer regions of the Galactic disc.

– What is the Galactic bulge? The Galactic bulge was for a long time considered to be a single population with fast formation and metal enrichment. However, recent studies have found that the bulge stars span a wide range of metallicities and ages. How much of the bulge could still be a classical bulge and not a buckled bar (pseudo-bulge)? What mass fraction in the bulge can be ascribed to halo stars? Can we explore the prominent X-shape of the bar and can we conclude — on the basis of ages and metallicities of the stars — when the buckling event happened?

### Science requirements

With 4MIDABLE-HR we aim to distinguish stellar populations in the elemental abundance plane over three quadrants of the Galactic disc, probing as deep as 10 kpc from the Sun. For this we need a sample size of more than two million stars with elemental abundances that are accurate to well within 0.1 dex. The sample size is necessary because a detailed understanding of the Milky Way requires resolving the full kinematic and chemical distributions in all stellar populations at different positions in the Galaxy. To characterise the moments of the kinematic distributions of a single population at a given location, one needs of order 100 stars to get a statistically robust mean down to 10% of sigma ($\sqrt{N}$). The precision in elemental abundances allows us to cut the sample in 0.1-dex width in abundance space. Sampling the abundance plane in [α/Fe] versus [Fe/H] meaningfully would thus require on the order of 50 abundance groups times 100 stars. The total sample size in this survey thus allows us to map out these populations at about 400 locations. This will enable us to distinguish the separation of elemental abundance trends between the α-poor and α-rich discs, and between the halo and the metal-poor disc, and to trace the position of the [α/Fe]–[Fe/H] "knee" from the solar neighbourhood, through the inner disc, and into the bulge (Hayden et al., 2011).

To resolve waves in the Galactic disc as seen by, for example, Antoja et al. (2018), one needs to resolve the mean velocity of the stars to better than 0.3 km s$^{-1}$, which, at the velocity dispersion of 50 km s$^{-1}$, requires 20 000 stars for each data point and would allow us to resolve about 100 different points in angular momentum space or physical position space. We also want to map the age-velocity dispersion relation at different points in elemental abundance space and physical position space to shed light on the heating mechanisms in the Galactic disc in order to differentiate between possible high velocity dispersion at birth for older stars, secular heating by disc structure (for example, the Galactic bar, or spiral arms) or giant molecular clouds, and merger induced heating. To detect a merger, we can roughly estimate that we need to precisely track an increase of 5 km s$^{-1}$ in the velocity dispersion over the course of roughly one billion years. This would require 5000 stars in each age bin or about 50 000 stars in total. These need to be sampled at 10 to 20 different Galactocentric radii and altitudes dissected in metallicity to pinpoint the radial pattern of heating. With these estimates, one would need on the order of 2.5 million stars to comprehensively address the formation of the Galactic components.

### Target selection and survey area

#### The Bulge and Disc field star sub-survey

The 4MIDABLE-HR selection function has been designed to be simple and reproducible. The magnitude limit is set to $G$ = 15.5 magnitudes, which allows us to obtain a high-quality spectrum with the required S/N = 100 per Å in a two-hour exposure. To avoid cool main sequence stars an upper limit on the absolute magnitude is set to $M_G$ < 5 magnitudes. Gaia parallaxes were used to estimate $M_G$. The range of sky declinations for 4MOST is set to $-80 < \text{dec} < +20$ degrees. We note that this area extends outside the fiducial survey footprint, which is restricted by prevailing northern winds, etc. (see Guglion et al., p. 17).

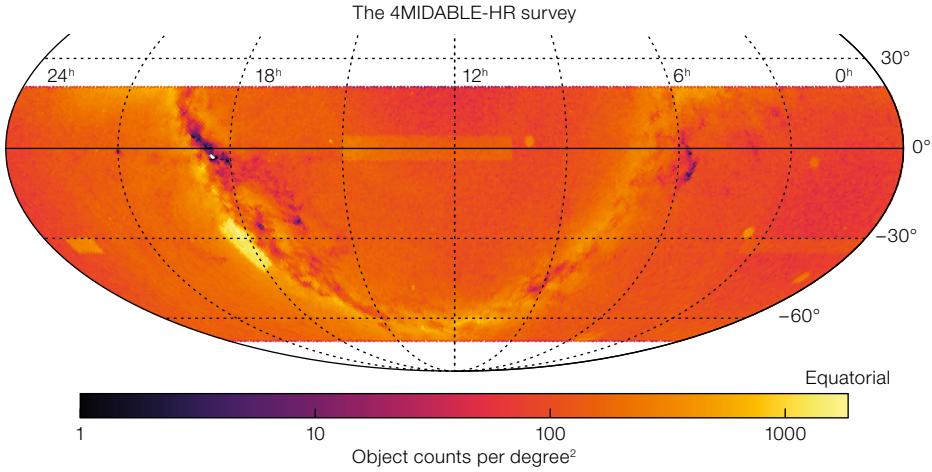

Figure 3. Density map of the input target catalogue. We aim to observe at least a random 50% of these targets. Note that the targets with dec > 5 degrees have a smaller likelihood of being observed in the fiducial survey strategy (Guiglion et al., p. 17).

Applying these simple cuts in dec, $G$, and $M_G$ provides us with a sample of more than 21 million targets from the Gaia Data Release 2. This sample is downsized to fit into a five-year survey, and as we need to have about twice as many targets in the catalogue to allow for an efficient usage of the fibres, our target catalogue contains about 4.1 million targets. Figure 2 shows the physical positions and the colour-magnitude diagram of the stars in this catalogue.

*Deep bulge fields*
The main bulge and disc sub-survey extensively probes the inner disc region and its transition into the bulge region. It is, however, not deep enough to provide a statistically significant number of targets within the central few kpc. Therefore, to better probe the properties of the central part of the Galaxy we designed a bulge deep field sub-survey with pointed observations down to $G = 17.5$ magnitudes in a grid pattern in the southern bulge between $-8 < l < 8$ degrees and $-10 < b < -4$ degrees. Up to eight hours will be spent in each of our 32 fields and the target catalogue contains about 72 000 targets.

*Deep disc fields*
To probe the vertical extent, the scale-height, and the interface between the disc and the halo, 4MIDABLE-HR will do deep observations in the footprints of the WAVES extragalactic survey that uses the low-resolution fibres. In the WAVES deep fields (about 65 square degrees) and in the WAVES wide fields (about 1300 square degrees) we will reach stars about ten times fainter and three times fainter, respectively, than the stars in our main disc and bulge catalogue. The catalogue contains about 65 000 stars. The positions of the WAVES fields can be found in Driver et al. (p. 46), but as stated therein, the exact location of the deep fields could be subject to change.

*Bulge Cepheid survey*
Classical Cepheids are young stars that trace the chemical composition of the interstellar medium. Type II Cepheids are post-horizontal-branch stars that trace the old population. They are present in the bulge, the thick disc, and probably the halo and allow 4MIDABLE-HR to study these Milky Way subsystems and their interfaces. The abundances of numerous elements can be derived from the analysis of Cepheid spectra. More importantly, their distances can be derived accurately (since they follow period-luminosity relations) even at large distances where Gaia parallaxes are less accurate. 4MIDABLE-HR will focus on Cepheids with $13.5 < G < 15.5$ magnitudes and our catalogue contains about 800 Cepheids towards the Galactic bulge.

### Spectral success criteria and the figure of merit

The desired precision of 0.05 dex is achievable in most elemental abundance ratios such as the $\alpha$-elements and most iron peak elements when S/N > 100 per Å is reached (based on simulations with the 4MOST Galactic Pipeline). We have not imposed any constraints on reddening in our target catalogue, as the results from the pipeline indicate that the impact is negligible for the stars we aim to observe. It is, however, important to keep in mind that some rare-earth elements that can only be measured in the blue spectral region, which will be difficult to obtain for highly reddened spectra.

The S/N requirement will be measured in the continuum in the wavelength range 6190–6210 Å, which is clean and free from strong spectral lines.

The figure of merit (FoM) is a measure of how successful the survey is, and for 4MIDABLE-HR, it is simply the ratio between the number of successfully observed targets and the 4.3 million stars in the input target catalogue. As long as stars are chosen and observed randomly from this catalogue, the survey will be regarded as successful when a FoM = 0.5 is reached.

### Acknowledgements

This work was supported by the project grant "The New Milky Way" from the Knut and Alice Wallenberg Foundation.